# Room temperature quantum Hall effect in a gated ferroelectric-graphene heterostructure

Anubhab Dey[1*], Nathan Cottam[1], Oleg Makarovskiy[1], Wenjing Yan[1], Vaidotas Mišeikis[2-3], Camilla Coletti[2-3], James Kerfoot[4#], Vladimir Korolkov[4#], Laurence Eaves[1], Jasper F. Linnartz[5], Arwin Kool[5], Steffen Wiedmann[5], Amalia Patanè[1*]

[1] School of Physics and Astronomy, University of Nottingham, Nottingham NG7 2RD, UK

[2] Center for Nanotechnology Innovation @NEST, Istituto Italiano di Tecnologia, Piazza San Silvestro 12, 56127 Pisa, Italy

[3] Graphene Labs, Istituto Italiano di Tecnologia, Via Morego, 30, 16163 Genova, Italy

[4] Park Systems UK Ltd , Medicity Nottingham, D6 Thane Road, Nottingham, NG90 6BH, UK

[5] High Field Magnet Laboratory (HFML –EMFL), Radboud University, Toernooiveld 7, 6525 ED Nijmegen, The Netherlands

*Author to whom any correspondence should be addressed.
Email : anubhab.dey@nottingham.ac.uk
E-mail: amalia.patane@nottingham.ac.uk
# Website www.parksystems.com

**Abstract** The quantum Hall effect is widely used for the investigation of fundamental phenomena, ranging from topological phases to composite fermions. In particular, the discovery of a room temperature resistance quantum in graphene is significant for compact resistance standards that can operate above cryogenic temperatures. However, this requires large magnetic fields that are accessible only in a few high magnetic field facilities. Here, we report on the quantum Hall effect in graphene encapsulated by the ferroelectric insulator $CuInP_2S_6$. Electrostatic gating of the graphene channel enables the Fermi energy to be tuned so that electrons in the localized states of the insulator are in equilibrium with the current-carrying, delocalized states of graphene. Due to the presence of strongly bound states in this hybrid system, a quantum Hall plateau can be achieved at room temperature in relatively modest magnetic fields. This phenomenon offers the prospect for the controlled manipulation of the quantum Hall effect at room temperature.

The electronic properties of graphene are very sensitive to applied magnetic fields (**B**) and are ideally suited for the investigation of the quantum Hall effect (QHE). This is exemplified by plateaus in the Hall resistance due to the quantization of the two-dimensional electron motion into Landau levels (LL)[1, 2, 3, 4, 5, 6, 7, 8, 9]. The QHE, first discovered in Si metal–oxide–semiconductor field-effect transistors [10], exhibits important differences in graphene due to the electron–hole degeneracy near the charge neutrality point, which leads to a distinctive half-integer QHE and a non-zero Berry's phase of the electron wavefunction [4, 5, 7, 8, 9].

Of particular significance for the QHE in graphene is the effect of dopant impurities near its surface. Screening effects in graphene [11, 12] tend to be weakened by a magnetic field and can facilitate the localisation of charge carriers in the disordered potential of the graphene layer [13, 14]. For example, for epitaxial graphene on a Si-terminated SiC substrate [15, 16, 17, 18], donors reside in the SiC layer adjacent to the graphene layer. These dopants act as a reservoir of electrons for graphene, maintaining the Hall voltage on the $v = 2$ QH plateau over a wide range of magnetic fields [19]. An extended quantum Hall plateau was also observed in graphene-based field effect transistors (FETs) in which graphene is capped by a thin layer of the van der Waals crystal InSe [20, 21]. These examples of "giant" QH plateaus in graphene have been reported at low temperatures ($T < 200$ K) and have been assigned to the magnetic field and electric field induced transfer of charge carriers between the degenerate Landau levels of graphene and the localized states in its proximity. A full microscopic model for the QHE in these hybrid systems does not yet exist. However, recent work has modelled the interaction between free carriers and localized charges near the surface of graphene [22], showing that when the chemical potential is in the gap between Landau levels, these charges can form stable bound states over a distance of the order of the magnetic length $l_B = \sqrt{\hbar/eB}$ and binding energy $E_B \approx (\hbar v_F/l_B)$, where $v_F \approx 10^6$ m/s is the Fermi velocity and $e$ is the elementary charge. This phenomenon can persist well beyond cryogenic temperatures, opening possibilities for the controlled manipulation of the QHE at room temperature. To date, a room temperature resistance quantum has been reported only in high mobility graphene at large magnetic fields that are available only in a few high field magnet laboratories [5, 23].

Here, we report on the QHE in field effect transistors based on single layer graphene capped with the ferroelectric van der Waals crystal $CuInP_2S_6$ (CIPS). The CIPS layer is used as a source of localized charge carriers in proximity to graphene. We report a hysteretic behaviour in the longitudinal and transverse magnetoresistance of graphene over a range of applied magnetic fields and temperatures. Similar hysteretic phenomena in the resistivity of graphene have been reported previously in zero magnetic field and assigned to charge trapping [24, 25, 26, 27, 28, 29] and/or ferroelectric polarisation [30, 31, 32, 33, 34]. In this work, we report on the dynamic exchange of charge carriers at the CIPS/graphene interface and its influence on the QHE and its hysteretic behaviour. The QHE is found to be weakly dependent on temperature and is observed at room temperature over a range of easily accessible applied magnetic fields.

## Results

**Transport characteristics in zero magnetic field** The CIPS/graphene heterostructure was prepared by exfoliation and visco-elastic stamping of a CIPS flake on a Hall bar based on high-quality graphene grown by CVD (chemical vapour deposition). Figure 1a shows the optical image of a ten-terminal Hall bar, half of which is based on graphene (G) and the other half on CIPS/graphene (CG), mounted on a 285 nm-thick $SiO_2$/$n$-Si substrate. The morphology of the layers was probed by atomic force microscopy (AFM) and single pass amplitude-modulated Kelvin probe force microscopy (AM-KPFM) [35]. The CIPS layer has a non-uniform thickness ranging from 20 nm to 50 nm and a uniform work function potential at the graphene/CIPS interface (Figure 1b). Details of the fabrication and of the characterisation of the CIPS flakes by AFM and piezoresponse force microscopy (PFM) are in the experimental section and Figure S1 of Supplementary Information (SI).

The longitudinal resistance $R_{XX}$ was measured at a constant current ($I = 1$ µA). The voltage drop $V_{XX}$ across different pairs of terminals along the graphene channel was measured over a range of gate voltages $V_G$ applied between the graphene and Si-gate electrodes. As can be seen in the inset of Figure 1c, for pristine graphene the $R_{XX}(V_G)$ curve at $T = 300$K is peaked at the neutrality point $V_{NP} = +10$ V. Using a capacitance model of the graphene FET, we estimate a hole density $p = 7 \times 10^{11}$ cm$^{-2}$ at $V_G = 0$ and a hole (electron) mobility $\mu = 9 \times 10^3$ cm$^2$/Vs ($1 \times 10^4$ cm$^2$/Vs) for carrier concentrations in the range $10^{11}$-$10^{12}$ cm$^{-2}$ at $T = 300$ K. In contrast to pristine graphene, for CG the $R_{XX}(V_G)$ curves show a pronounced hysteresis and are asymmetric (Figure 1c): The amplitude of the hysteresis increases with increasing the sweep range of $V_G$ from $\Delta V_G = \pm 10$ V to $\pm 50$ V. For $\Delta V_G = \pm 10$ V, the $R_{XX}(V_G)$ curves are shifted to lower values of $V_G$ compared to pristine graphene; also, the field effect mobility (and Hall mobility) for holes and electrons is reduced from $\mu \sim 10^4$ cm$^2$/Vs to $\mu \sim 2 \times 10^3$ cm$^2$/Vs. In general, the $R_{XX}(V_G)$ curve consists of multiple peaks, suggestive of a channel with a non-uniform distribution of dopants; also, the temporal response of $R_{XX}$ is slow (with rise and decay times $\tau > 100$ s). Thus, the $R_{XX}(V_G)$ curve depends on the sweep range of $V_G$ and sweep rate $\Delta V_G/\Delta t$. A value of $\Delta V_G/\Delta t = 0.3$V/s was used for the data presented in this work.

Hysteresis in the transport characteristics of graphene can arise from a gate-induced polarization at the interface of graphene with a ferroelectric layer [33]. For our CG, the hysteresis is not dominated by this phenomenon as a gate-induced ferroelectricity would produce a shift of the neutrality point $V_{NP}$ in the direction of the gate sweep, *i.e.* $V_{NP}$ would shift to higher voltages when $V_G$ is swept from negative to positive values compared to when $V_G$ is swept from positive to negative values. On the other hand, a hysteresis can also originate from a slow charge transfer at the CIPS/graphene interface, as reported earlier in a similar device structure[25]. The gate voltage induces charges in the graphene layer, which then redistribute between the graphene and CIPS layers. In the first part of the sweep of $V_G$ to positive gate voltages ($V_G > 0$ V), electrons are transferred from graphene onto CIPS; during the reverse sweep with $V_G < 0$ V, the CIPS layer discharges its electrons onto graphene. A non-homogeneous distribution of localized states in CIPS can create areas of graphene with different carrier densities, thus causing the multiple peaks in $R_{XX}(V_G)$ shown in Figure 1c.

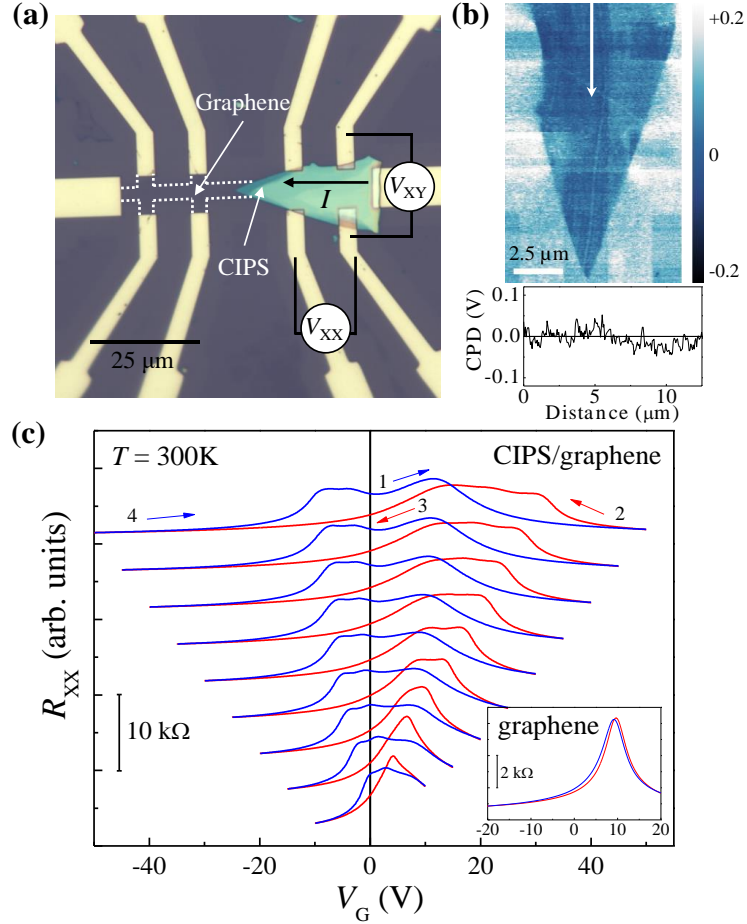

**Figure 1: Gated Hall bar based on graphene capped with CIPS** (a) Optical image of a Hall bar based on CIPS/graphene (CG) on a SiO$_2$/$n$-Si substrate and Ni-Au contacts. One section of the graphene layer is covered by a CIPS layer. The white dotted lines mark the edges of pristine graphene. (b) AM-KPFM contact potential difference (CPD) map (top) and CPD profile (bottom) of CG measured with a Multi75E cantilever at a voltage amplitude of $V_{AC} = 4$ V and frequency $f_{AC} = 17$ kHz. The CPD-profile is obtained along the length of the CIPS-flake, as indicated by the white arrow in the map. (c) Resistance-gate voltage $R_{XX}(V_G)$ curves for CG at $T = 300$ K ($I = 1$ µA, $B = 0$ T). The sweep up/down branches are shown in blue and red arrows, respectively. Curves are displaced along the vertical axis for clarity. Inset: $R_{XX}(V_G)$ curve for a reference sample based on pristine graphene (G) at $T = 300$ K ($I = 1$ µA, $B = 0$ T). This sample corresponds to the uncapped section of the graphene Hall bar shown in part (a). A sweep rate $\Delta V_G/\Delta t = 0.3$ V/s was used for the measurements.

We model the hysteresis in $R_{XX}(V_G)$ using a classical capacitance model of the FET that takes into account a charge transfer at the CIPS/graphene interface (Figure S2 in SI). We estimate that a charge $\Delta Q$ redistributes slowly between the graphene ($Q_g$) and CIPS ($Q_{CIPS}$) layers with a characteristic time constant $\tau > 100$ s; different regions of CG tend to charge/discharge with similar temporal dynamics; also, the value of $\Delta Q/e = n_Q$ is dependent on $V_G$ and reaches values of up to $n_Q \sim 10^{12}$ cm$^{-2}$ at large $V_G$ ($V_G = +50$ V) and $T = 300$ K.

The hysteresis in $R_{XX}(V_G)$ weakens with decreasing $T$ (Figure 2(a)) or under excitation of the sample with photons of energy larger ($h\nu = 3.06$ eV) than the band gap of CIPS (Figure S3). Light of increasing intensity induces a shift of the neutrality point to larger positive $V_G$ and a narrowing of the $R_{XX}(V_G)$ curve. This indicates that carriers photocreated in the CIPS layer can screen the disordered potential created by localized charges. In summary, the transport characteristics of graphene are very sensitive to charges trapped in the CIPS layer. This effect is observed in all our CG devices and is used to probe the effect of localized charges on the QHE at different temperatures, magnetic fields and gate voltages.

**Magneto-transport and quantum Hall effect** Figures 2a and 2b show the temperature dependence of the $R_{XX}(V_G)$ curves for CG at $B = 0$ T and 16 T respectively. At low temperatures ($T \leq 200$ K), the hysteresis in $R_{XX}(V_G)$ is weak, as also observed in the pristine graphene. However, it becomes pronounced for $T > 200$ K. In particular, in a magnetic field ($B = 16$ T in Figure 2b), the $R_{XX}(V_G)$ curves exhibit additional maxima and minima. To illustrate this behaviour more clearly, we plot in Figure 2c-d-e the colour maps of $R_{XX}$ versus $V_G$ and $B$ at different $T$ and for different (up/down) sweeps of $V_G$. For $T$ up to 200 K (Figure 2c-d), the bright red region in $R_{XX}(B, V_G)$ centred at $V_{NP} \sim +30$ V corresponds to the neutrality point of graphene represented by the zeroth Landau level, LL ($n = 0$). For both sweep up/down branches, secondary peaks in $R_{XX}(B, V_G)$ emerge for $B > 5$ T at around $V_G = +20$ V and $+40$ V. As $V_G$ increases from negative to positive values, first holes ($V_G < V_{NP}$) and then electrons ($V_G > V_{NP}$) fill successive LLs.

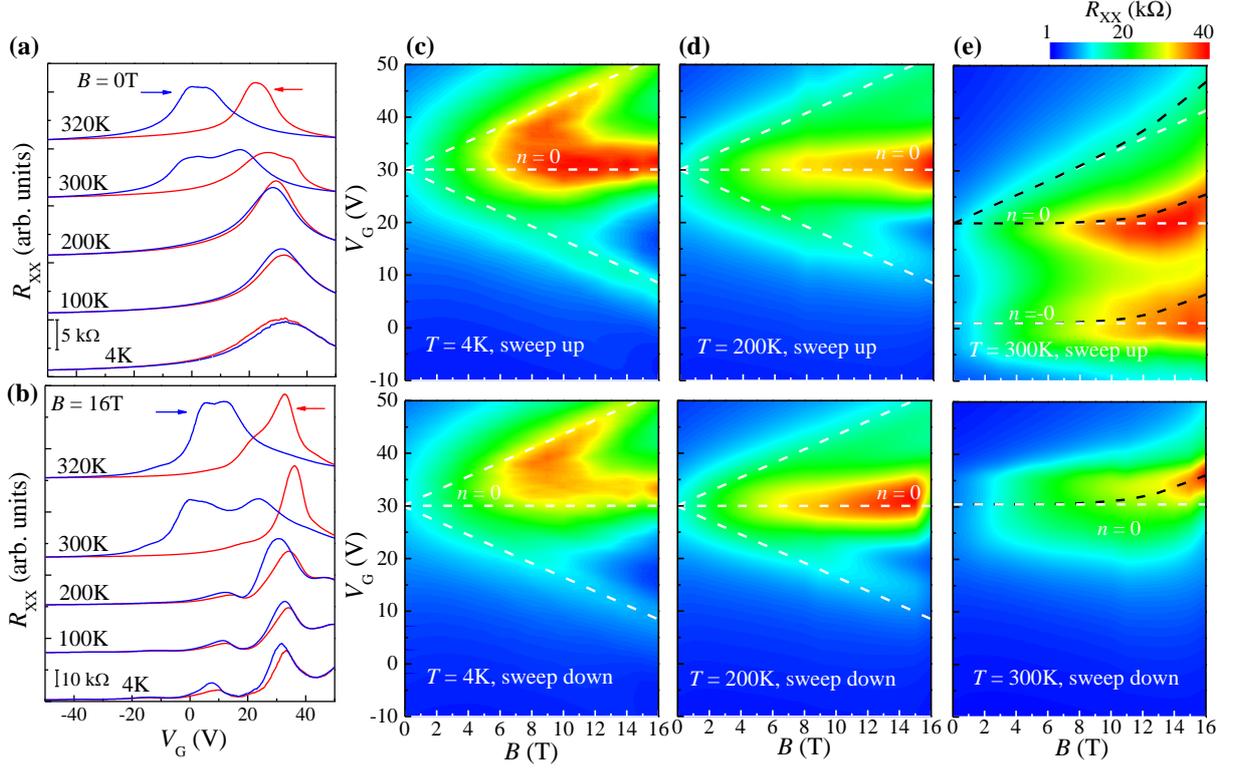

**Figure 2: Longitudinal magnetoresistance for CIPS/graphene** (a-b) Resistance-gate voltage $R_{XX}(V_G)$ curves for CIPS/graphene at different temperatures $T$ ($I = 1$ µA) and for (a) $B = 0$ T and (b) $B = 16$ T. The sweep up/down branches are shown in blue and red, respectively. For clarity, curves are displaced along the vertical axis. (c-d-e) Colour plots of $R_{XX}$ versus $B$ and $V_G$ at (c) $T = 4$ K, (d) $T = 200$ K and (e) $T = 300$ K and different sweeps (top: sweep up; bottom: sweep down; $I = 1$ µA). Dashed white lines represent the calculated Landau level (LL) charts using a conventional model, as described in the text. Dashed black lines in part (e) show the calculated LL charts assuming a $B$-dependent charge transfer.

The energy-level spectrum of Dirac fermions in a magnetic field is described by the relation $E_n = sgn(i)\sqrt{2e\hbar v_F^2 B |n|}$, where $n = 0, \pm 1, \pm 2...$ The spectrum comprises electron and hole LLs, as well as a LL ($n = 0$) at the neutrality point. We use the capacitance equation $C = e[dn_g/dV_G]$ to calculate the voltage separation $\Delta V_G$ of the maxima in the $R_{xx}(V_G)$ curve at different $B$. Here $C = \varepsilon\varepsilon_0/t$ is the "classical" capacitance per unit area of the graphene/SiO$_2$/Si heterostructure, $t = 285$ nm is the SiO$_2$ layer thickness, $\varepsilon = 3.9$ is the relative dielectric constant of SiO$_2$, $\varepsilon_0$ is the permittivity of free space, and $n_g$ is the carrier density in the graphene layer. We express the separation between the two maxima in $R_{xx}(V_G)$ corresponding to the alignment of the Fermi level with the $n = 0$ and $n = \pm 1$ LLs as $\Delta V_G = eg/C$, where $g = 4eB/h$. This model reproduces the data at low $T$ for both sweep up and down of $V_G$ ($T = 4$ K and 200 K in Figure 2c-d, white dashed lines), but fails to describe the data at $T > 200$ K ($T = 300$ K in Figure 2e, white dashed lines). At $T = 300$ K, the LL quantization is obscured by a large hysteresis; in particular, the neutrality point $V_{NP}$ shifts to larger positive $V_G$ with increasing $B$. The black lines in Figure 2e describe the deviation of the LL features in $R_{xx}(B, V_G)$ from a conventional LL chart model. The measured deviation is reproduced by considering a $B$-dependent charge transfer and the capacitance equation $C = e[dn_g/dV_G]$. The magnetic field tends to reduce the density of electrons transferred from CIPS to graphene by $\Delta n_g = 4 \times 10^{10}$ cm$^{-2}$ at $B = 10$ T and $\Delta n_g = 4 \times 10^{11}$ cm$^{-2}$ at $B = 16$ T. This phenomenon can also be seen in the dependence of the Hall resistance ($R_{XY}$) on $B$, $V_G$ and $T$, as discussed below.

Figure 3 shows the $V_G$-dependence of $R_{XY}$ over a range of temperatures ($T = 4$-320 K) and magnetic fields from $B = 0$ T to 16 T. At $T = 4$ K (Figure 3a), the $R_{XY}(V_G)$ curve shows QH plateaus centred at $V_G \approx +20$ V and $V_G \approx +40$ V, corresponding to the LL filling factor $v = 2$ for holes and electrons, respectively. The LL filling factor $v$ is derived from the relation $v = \pm 4(|n|+1/2)$, where $n$ is the LL index [16]. Plateaus corresponding to lower value of $R_{XY}$ can also be seen at $V_G \approx -2$ V ($v = 6$) and $V_G \approx -25$ V ($v = 10$). As the temperature increases to $T = 100$ K (Figure 3b) and 200 K (Figure 3c), the QH plateaus tend to narrow. A further increase of temperature to $T \geq 300$ K (Figure 3d and 3e) induces a prononuced hysteresis in the $R_{XY}(V_G)$ curves (see also Figure S4 and S5 in SI). Figure 4 shows the colour plots of $R_{XY}$ versus $V_G$ and $B$ at $T = 300$ K for different (up/down) sweeps of $V_G$. These data illustrate the sign of the $v = 2$ QH plateau and its evolution with increasing values of $B$ and $V_G$. In particular, it can be seen that the neutrality point $V_{NP}$ shifts to larger positive $V_G$ with increasing $B$.

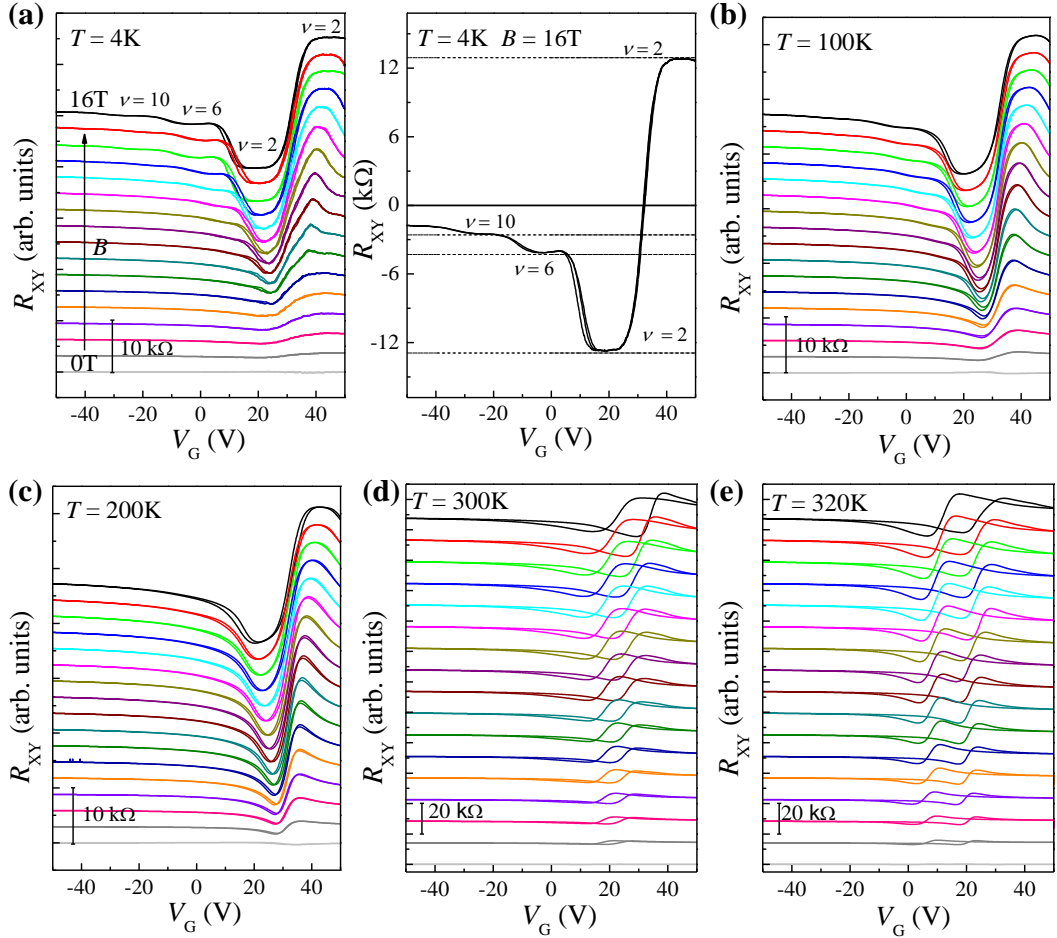

**Figure 3: Hall resistance for CIPS/graphene** (a-b-c-d-e) Hall resistance-gate voltage $R_{XY}(V_G)$ curves at different temperatures: (a) $T = 4$ K, (b) $T = 100$ K, (c) $T = 200$ K, (d) $T = 300$ K and (e) $T = 320$ K, and magnetic field ranging from $B = 0$ T to 16 T in 1 T steps ($I = 1$ μA). The right panel in part (a) shows the $R_{XY}(V_G)$ curves at $B = 16$T and $T = 4$K. Dashed lines correspond to the quantized values of $R_{XY}$.

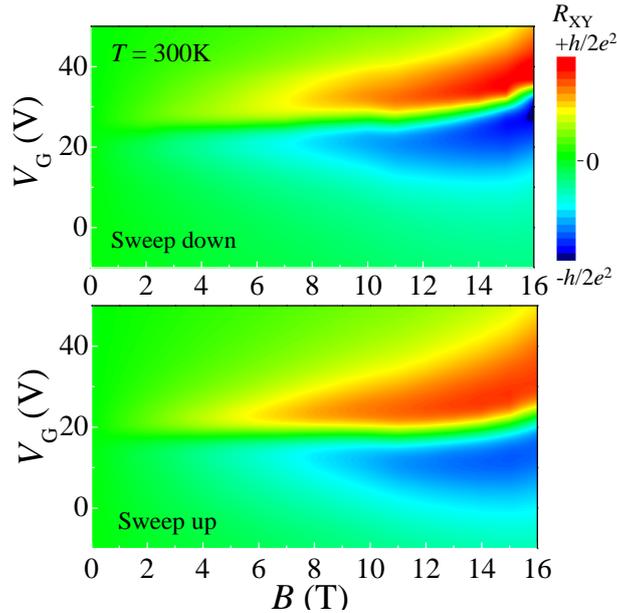

**Figure 4: Room temperature quantum Hall resistance for CIPS/graphene** Colour plots of the Hall resistance $R_{XY}$ versus magnetic field $B$ and gate voltage $V_G$ at $T = 300$ K ($I = 1$ μA) for different sweeps of $V_G$ (top: sweep down; bottom: sweep up).
999

From Figure 3 it can be seen that the $v = 2$ QH plateau is accompanied by a hysteresis that depends on $T$ and $B$. This behaviour is shown in more detail in Figure 5a where the $R_{XY}(V_G)$ curves are plotted at $B = 16$ T for different $T$. To quantify the hysteresis, we consider the gate voltage at which $R_{XY}(V_G)$ goes to zero (*i.e.* the charge neutrality point) on the sweep up ($V_{Gu}$) and sweep down ($V_{Gd}$) branches of $R_{XY}(V_G)$. The difference between the two values, $|\Delta V_G| = |V_{Gu} - V_{Gd}|$, is shown in Figure 5b for different $T$ and $B$. For $T < 200$ K, $|\Delta V_G|$ is weakly dependent on $T$ and tends to increase with $B$. For $T > 200$ K, the hysteresis is more pronounced and can be described by the relation $|\Delta V_G| \propto \exp(-E_a/kT)$, where $E_a$ is an activation energy given by $E_a \approx 0.16$ eV for $B = 16$ T (Arrhenius plot in the inset of Figure 5b). Figure 5a also reveals that increasing $T$ above $T = 100$ K leads to a shift of the neutrality point to lower values of $V_G$, corresponding to an increasing density of electrons in the graphene layer. This behaviour is not observed in pristine graphene and is assigned to the thermal excitation of electrons from CIPS into the graphene layer. For lower $T$ ($T = 4.2$ and 100 K), the shift of the neutrality point is towards higher values of $V_G$ with increasing $T$, indicative of a thermal excitation of carriers near the Dirac point.

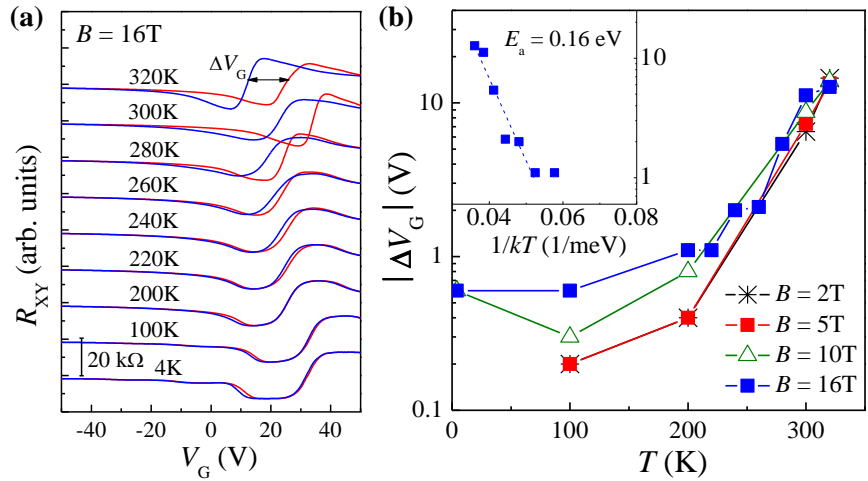

**Figure 5: Hysteresis in the Hall resistance of CIPS/graphene** (a) $R_{XY}(V_G)$ at $B = 16$ T and different temperatures $T$. The sweep up/down branches are shown in blue and red, respectively. For clarity, curves are displaced along the vertical axis ($I = 1$ μA). $\Delta V_G$ is the amplitude of the hysteresis in $R_{XY}(V_G)$, as estimated from the voltage at which $R_{XY}(V_G) = 0$. (b) Amplitude of the hysteresis $|\Delta V_G|$ versus $T$ at different $B$. For $T = 4.2$ K and $B \leq 5$ T, $|\Delta V_G| \approx 0$. Inset: Arrhenius plot of $|\Delta V_G|$ versus $1/T$ at $B = 16$ T. The dashed line is an exponential fit to the data.

Due to the hysteresis and slow charge transfer in CG, the measurement of $R_{XX}$ and $R_{XY}$ versus $B$ at a given $V_G$ require special consideration. For each measurement of the $R_{XY}(B)$ and $R_{XX}(B)$ curves, the value of $V_G$ was increased by small increments ($\Delta V_G/\Delta t = 0.1$ V/s) starting from $V_G = 0$ V until reaching the required value of $V_G$. The temporal dependence of $R_{XY}$ and $R_{XX}$ at $B = 0$ T was then followed over intervals of several minutes, as required for $R_{XY}$ and $R_{XX}$ to reach stable values. The magnetic field was then swept from $B = 0$ T to 16 T (sweep rate of 5 mT/s). The values of $V_G$ were selected according to the $R_{XY}(V_G)$ curves in Figures 3 and 4, showing plateaus on each side of the neutrality point (between the $n = 0$ and $n = \pm 1$ LLs) due to holes ($V_G \approx +20$ V) or electrons ($V_G \approx +40$ V).

Figure 6a shows the $R_{XY}(B)$ curves of CG for $V_G = +20$ V at $T = 4$ K and 300 K. It can be seen that the $R_{XY}(B)$ curves exhibit a weak $T$-dependence; in particular, the approach to the $v = 2$ QH plateau shifts to lower $B$-fields at $T = 300$K. The value of $R_{XY}$ at the plateau and its stability over time depend on the gate voltage. We have observed similar behaviours in other devices (Figure S8-S9-S10), although the threshold in $B$ for the $v = 2$ QH plateau may differ depending on the quality of the graphene layer, which can contain defects and impurities introduced during the growth and/or the transfer of CVD-grown graphene from the residual Cu onto the $SiO_2$/Si substrate. The behaviour of CIPS/graphene contrasts with the strong temperature dependence of the $v = 2$ QH plateau in pristine graphene (Figures 6b and S11). Although the analysis of the QH plateau in the proximity of the charge neutrality point is complicated by the contribution of both electrons and holes to the conductivity, we can select gate voltages at which a QH plateau is observed for both holes and electrons (Figures 4, 6c-d).

The plateau in $R_{XY}(B)$ is accompanied by a corresponding decrease in $R_{XX}(B)$ (Figure 6e and S6-S7). However, we note that $R_{XX}$ does not go to zero at values of $B$ corresponding to the $v = 2$ QH plateau; also, we observe a small deviation of $R_{XY}$ from its nominal quantized value ($h/2e^2$). This can also be seen in pristine graphene at low $T$ (Figure S11). As shown in Figure 6f, this deviation ($\Delta R_{XY}$) depends on $R_{XX}$ and tends to zero for decreasing $R_{XX}$. Here, values of $R_{XX}$ and $R_{XY}$ are obtained from measurements of the same device at different $T$, $B$ and/or $V_G$ after the onset of the $v = 2$ QH plateau in $R_{XY}(B)$.

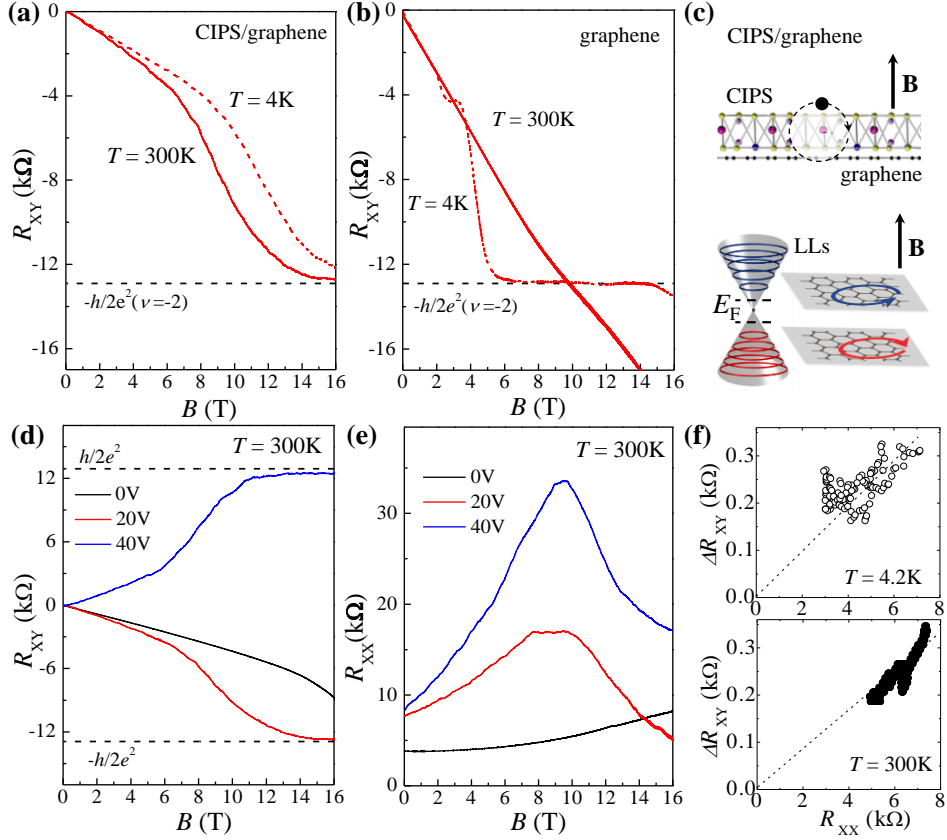

**Figure 6: Quantum Hall plateau in CIPS/graphene (CG)** (a-b) Hall resistance $R_{XY}$ versus magnetic field $B$ at $T = 4$ K and 300 K in (a) CG ($V_G = +20$ V, $I = 1$ µA) and (b) pristine graphene ($V_G = +3$ V, $I = 1$ µA). (c) Top: Schematic of bound states in CIPS/graphene. Bottom: Landau levels (LLs) in graphene with Fermi level aligned between the $n = 0$ and $n = \pm 1$ LLs corresponding to the $v = 2$ QH plateau. (d) $R_{XY}$ versus $B$ for CG at different $V_G$ and $T = 300$ K ($I = 1$ µA). Negative and positive values of $R_{XY}$ refer to hole and electron resistivity, respectively. (e) $R_{XX}$ versus $B$ for CG at different $V_G$ and $T = 300$ K ($I = 1$ µA). (f) Deviation of $R_{XY}$ from the quantized value ($R_{XY} = h/2e^2$) versus $R_{XX}$, as derived from measurements at different $T$, $B$ and $V_G$ (top: $T = 4.2$ K, $B = 16$ T, $V_G = 42$-$28$ V and $16$-$22$ V; bottom: $T = 300$ K, $B = 14$-$16$ T, $V_G = 20$ V). Dashed lines are guides to the eye.

Our data indicate a coupling between $R_{XX}$ and $R_{XY}$ that could be accounted for by geometrical effects and/or disorder. A non-uniform channel can exhibit regions that do not have minimal resistance at the same magnetic field. This can result in an effective misalignment of the Hall probes so that the measured Hall resistance $R_{XY}$ is influenced by the longitudinal resistance $R_{XX}$ [36]. Also, disorder can play an important role on the ν = 2 QH plateau due to the coexistence and contribution to the transport of both electrons and holes[37]. Using the data in Fig. 6f, we estimate the coupling parameter $s = \Delta R_{XY}/R_{XX}$. A linear fitting of $\Delta R_{XY}$ versus $R_{XX}$ indicates $s$ = 0.04 (0.05) at $T$ = 300K ($T$ = 4.2K). Our values are similar to the value ($s$ = 0.038) for high-quality graphene/SiC devices reported in the recent literature [38].

## Discussion

The room temperature QHE was first reported in graphene and explained in terms of the magnetic field quantization of Dirac fermions in graphene [5]. The LL quantization energy of fermions in magnetic field is $E_n = v_F \sqrt{2e\hbar B |n|}$. For $n = \pm 1$ and $B$ = 45 T, $E_n \sim$ 250 meV, which greatly exceeds the thermal energy (~ 26 meV) of charge carriers at $T$ = 300 K. However, the physics of the QHE in graphene is more complex and requires an understanding of the unique nature of the $n$ = 0 LL [39]. The measured thermal activation energy for the quenching of the ν = 2 QH plateau in graphene approaches the cyclotron energy gap only at high magnetic fields ($B \approx$ 30 T). At high $B$, the number of states with zero energy ($n$ = 0 LL) is determined by the total magnetic flux and does not depend on disorder. Thus, the $n$ = 0 LL is well separated from its neighbouring ($n = \pm 1$) LLs and the activation energy corresponds approximatively to $E_1 = v_F \sqrt{2e\hbar B}$; however, for lower $B$, LL mixing due to disorder broadens the LLs by means of inter-LL scattering, leading to an activation energy that is smaller than $E_1$. Thus, the observation of the ν = 2 QH plateau at room temperature in graphene requires a large $B$ accessible only in a few high field facilities. In our CIPS/graphene sample the ν = 2 QH plateau is observed at relatively small $B$ at $T$ = 300 K, yet it is not seen in pristine graphene in the same range of $B$; thus, our results merit further consideration.

First, we note that the CIPS layer can act as a remote source of carriers for graphene. Our measurements demonstrate that the charge transfer at the CIPS/graphene interface is tuneable by gating and is temperature dependent. Regions of CIPS with different densities of localized states tend to charge and discharge with a similar slow (~ 100 s) time constant at room temperature, thus accounting for the gate-induced hysteresis in the transport characteristics (Figure 1c). The hysteresis becomes significant only at $T > 200$ K (Figure 2), symptomatic of thermal activation and slow transfer of charges from/to the CIPS layer onto graphene.

At sufficiently high temperatures ($T > 200$ K), electrons (and holes) in the localized states of CIPS are in equilibrium with the current-carrying, delocalized states of graphene. Under these conditions, bound states are formed in CIPS/graphene. In contrast, at low $T$ such equilibrium cannot be established and the hysteretic behaviour is not observed. Also, a comparison of the transfer characteristics and their hysteresis for CG at $B = 0$ and 16 T indicates that the charge transfer is influenced by magnetic field. The magnetic field acts to enhance the hysteretic behaviour and distorts the $V_G$-dependence of $R_{XX}$ and $R_{XY}$ around the charge neutrality point. As shown in Figure 2e and 4, for $B \geq 10$ T the colour regions in $R_{XX}$ and $R_{XY}$ corresponding to the zeroth LL tend to shift to larger $V_G$ with increasing $B$, consistent with a reduced transfer of electrons from the CIPS layer onto graphene due to an increased localization of charges in the quantizing magnetic field. This can also be seen in Figure 5b, where the hysteresis (as measured by $|\Delta V_G|$) increases with $B$.

Reference 22 offers an insight into the role of localized charges near the surface of graphene: For a range of chemical potentials inside the gap between the zeroth and first LLs, charged impurities can form stable "molecules" bound by free carriers of opposite sign within graphene [22]. The optimal distance between charges in the bound state is of the order of the magnetic length $l_B = \sqrt{\hbar/eB}$ and their binding energy scales as $E_B = (\hbar v_F / l_B)$. For $B = 16$ T, this gives $l_B = 6.4$ nm and $E_B = 0.10$ eV. This binding energy is comparable to our estimate (0.16 eV) derived from the $T$-dependent hysteresis in $R_{XY}$ (Figure 5b). Thus, for sufficiently high $T$, electrons (and holes) in the localized states have a binding energy that exceeds the cyclotron energy gap.

We note that a strongly disordered system cannot show the QHE because no LL quantization can occur. However, it is well established that the standard picture of the QHE requires the existence of disorder and localized states. This enables the Fermi level to be pinned at energies between the extended states of adjacent LLs. Disorder is a key feature of the QHE and its thermal stability: it acts to pin the Fermi level between the LLs and maintains the Hall voltage on the plateaus. The charge transfer between the CIPS and graphene layers is reversible, leading to the $v = 2$ QH plateau for both electrons and holes, as shown in Figure 4 and 6.

We now consider our findings in the context of ongoing research on other hybrid systems based on graphene. For example, the use of a conducting layer, such as the relatively small band gap semiconductor InSe (~ 1.3 eV at $T = 300$ K) to form an InSe/graphene FET, facilitates the observation of a giant QH plateau, but its observation at room temperature is prevented by parallel conduction in the InSe layer [20, 21]. The use of a high-resistance dielectric poses other challenges. A giant QH plateau has been reported in graphene grown epitaxially by thermal annealing of a SiC dielectric substrate. In this case, the charge transfer at the SiC/graphene interface involves defects in SiC with a high densities of states ($10^{14}$ cm$^{-2}$ eV$^{-1}$) in close proximity to graphene [15, 16, 17, 18, 19]. These states arise from atomic-scale defects within the top few SiC layers, which are created during the formation of the graphene layer by Si-sublimation. For graphene on SiC, the giant QH effect was reported at temperatures of up to $T$ ~ 100-200 K, suggesting that the bound states in SiC/graphene have a relatively small binding energy even at high $B$ ($B > 20$ T). Alternatively, hexagonal boron nitride (hBN) represents an ideal dielectric for graphene-based FETs [40]. Charge and surface fluctuations in hBN tend to be weaker than in other substrates, such as SiO$_2$. Thus, graphene on hBN has a high-mobility and is well suited for observations of integer and fractional QHE [41, 42]. In particular, the formation of moiré superlattices in rotationally misaligned graphene/hBN layers can promote interfacial charge transfer and new quantum transport regimes [43, 44, 45, 46]. More recently, a hybrid system based on CrOCl-graphene also revealed an exotic QH effect phase due to the formation of a long-wavelength charge ordering [47]. In all these different hybrid systems, the band structure of graphene is modified around the Dirac cone as a result of an interfacial charge transfer involving a semiconductor or an insulator. However, for all these systems the observation of quantum effects at high temperatures, approaching room temperature, has proven to be challenging. Our choice of CIPS provides an effective layer for charge transfer as CIPS is a dielectric and its defect states are not only sufficiently dense ($\sim 10^{12}\,\text{cm}^{-2}\,\text{eV}^{-1}$), but also they form bound states with graphene that are sufficiently deep to be resilient to ionization at room temperature. The range of high temperatures ($T > 200$ K) for charge transfer and hysteresis in the transport curves corresponds to that required for activating the thermal motion of the Cu-ions [48, 49] out of the CIPS layer planes. This can lead to localized ionic charges whose slow motion could be responsible for the slow dynamics of charge transfer at the graphene/CIPS interface, leading to the hysteretic transport observed in this system. Since CIPS is a dielectric layer and electrons remain bound onto its localized states, the QH voltage in graphene is not short-circuited by a significant parallel conduction in the CIPS layer.

In conclusion, the controlled transfer of charges between graphene and localized states in its proximity provide a route for the observation of quantum effects at room temperature and in readily accessible magnetic fields. We have shown that the electric field-induced transfer of charge between the localized states in the CIPS and graphene layers acts to increase or decrease the carrier density in graphene, causing a change in its resistance that is gate-tuneable at high temperatures ($T > 200$ K). The charge transfer causes hysteretic behaviour in the electrical characteristics due a slow dynamic exchange of electrons between graphene and localized states in its proximity. Prospects for further research include a more accurate resistance quantization, which will require progress in both material growth and fabrication processes. This requires high-mobility homogenous graphene, a homogenous charge transfer at the graphene/CIPS interface, and the fabrication of low-resistance contacts. A more uniform

CIPS/graphene heterostructure could be achieved by the development of scalable growth techniques (for example using epitaxial graphene grown with intrinsic structural alignment on SiC) together with the fabrication of high-quality electrical contacts, such as electrodes with the edge-contact geometry [50]. Thus, there are prospects for further studies and for engineering interfacial charge transfer in hybrid systems based on graphene for the observation of quantum effects over a wide parameter space beyond the current state-of-the-art for future applications, such as graphene-based resistance standards for the new International System of Units [51].

## Methods

**Materials and device fabrication.** The $CuInP_2S_6$ crystal was purchased from HQ Graphene. Graphene Hall bars were fabricated at the NEST laboratories at the Istituto Italiano di Tecnologia, Pisa, Italy. The fabrication of high-mobility CVD-grown graphene Hall bars before the deposition of CIPS is crucial for the observation of the quantum Hall effect. Single-crystal graphene used in this work was grown on Cu foil by CVD in a cold-wall reactor (Aixtron BM Pro) using chromium nucleation seeds [52]. Graphene crystals were electrochemically delaminated from the growth substrate in 1M NaOH and deposited on $SiO_2$/n-doped Si wafers using semi-dry transfer [53]. The fabrication of the Hall bars was carried out using e-beam lithography (20 kV, Raith Multibeam on Zeiss Ultra Plus scanning electron microscope). Graphene Hall bars were prepared using reactive ion etching (gas flow 80 sccm $O_2$ and 5 sccm Ar, RF power 35 W) and electrical contacts were deposited by thermal evaporation of 7 nm of Ni and 60 nm of Au. A Poly(methyl methacrylate) (PMMA) resist (950 K, 4.5% in anisole, Allresist) was used for lithography, followed by 2-step cleaning in acetone and removal of AR600-71 (Allresist) to ensure a polymer-free surface of graphene [54]. The wafer containing the graphene Hall bar devices was covered with a protective coating of PMMA for dicing and storage. The processed wafers were diced into ~ 4x4 $mm^2$ chips and then cleaned in hot acetone ($T \sim 65$ °C) for 1 h, rinsed with isopropyl alcohol (IPA) and dried with pressurised nitrogen gas to remove the protective PMMA coating. These devices were then annealed in a tube furnace at $T = 300$ °C for 3 h in a 5% $H_2$ and 95% Ar flowing gas atmosphere to remove surface impurities and residues on the graphene surface. The graphene was then used for stamping the CIPS layer to form the CIPS/graphene heterostructure. The interface between graphene and CIPS after stamping was not further cleaned. The heterostructure was fabricated by exfoliating a CIPS flake onto polydimethylsiloxane (PDMS) from a low-residue tape and identified using optical microscopy. By using a micromanipulator stage, an exfoliated flake of CIPS on PDMS was aligned to one section of the graphene Hall bar and brought into contact with it. The PDMS was then slowly retracted in order to deposit CIPS. The graphene Hall bar capped with CIPS was then bonded into non-magnetic chip carriers for electrical measurements.

**Optical, electrical and microscopy studies**. The surface topography of the flakes was acquired by atomic force microscopy (AFM, Park NX20) in non-contact mode under ambient conditions. The KPFM study was conducted using an additional lock-in amplifier connected to the same AFM system. Transport measurements in the dark and under light illumination were conducted in vacuum ($2\times10^{-6}$ mbar) using Keithley-2400 source-meters and Keithley-2010 multi-meters. A temperature controller from Lakeshore Cryotronics was used to control and probe the temperature. A solid-state laser ($\lambda = 405$ nm) and a He-Ne laser ($\lambda = 632.8$ nm) were used for the optical studies. The position of the laser spot was adjusted on the device and measurements were taken at different powers. A cryogen free magnet (Cryogenic Limited) was used to perform the magneto-transport studies over a range of temperatures.

**Acknowledgements**

This work was supported by the European Union's Horizon 2020 research and innovation programme Graphene Flagship Core 3; the Engineering and Physical Sciences Research Council (Grant No. EP/M012700/1) and the University of Nottingham Propulsion Futures Beacon. Measurements in high magnetic field were supported by the European Magnetic Field Laboratory (EMFL) and by the EPSRC via the UK membership of the EMFL (Grant No. EP/X020304/1).

**Author contributions**

A.P. and A.D. conceived the project and wrote the paper; C.C and V.M. fabricated the graphene Hall bars; A.D. fabricated the CIPS/graphene devices and performed the transport studies assisted by N.C., O.M. and A.P.; A.D. conducted the analysis of the data, assisted by O.M. and A.P.; J.K. and V.K. conducted the microscopy studies; W.Y. contributed to the transport studies of CIPS in zero magnetic field; C.C. and V.M. synthesized and transferred single-crystal CVD graphene and fabricated the Hall bars; J.F.L. and S.R.W. contributed to the transport studies in high magnetic fields; all authors discussed the results.

**Additional information**

Competing financial interests: The authors declare no competing financial interests.

**Data availability**

The data that support the findings of this study are available from the corresponding author upon reasonable request.